\documentclass[10pt]{article}
\usepackage[T1]{fontenc}
\usepackage[a4paper]{geometry}
\usepackage{amssymb}
\usepackage{graphicx,psfig}

\newcommand{\bm}[1]{\mbox{\boldmath $ #1 $}}
\def\labelitemi{$\bullet$}

\newtheorem{thm}{Theorem}[section]
\newtheorem{lem}[thm]{Lemma}
\newtheorem{cor}[thm]{Corollary}
\newtheorem{prop}[thm]{Proposition}
\newtheorem{rem}[thm]{Remark}
\newtheorem{defi}[thm]{Definition}
\newtheorem{conj}[thm]{Conjecture}
\newcommand{\bconj}{\begin{conj}}
\newcommand{\econj}{\end{conj}}
\newcommand{\bth}{\begin{thm}}
\newcommand{\ethGL}{\end{thm}}
\newcommand{\bl}{\begin{lem}}
\newcommand{\el}{\end{lem}}
\newcommand{\bdf}{\begin{defi}}
\newcommand{\edf}{\end{defi}}
\newcommand{\bcor}{\begin{cor}}
\newcommand{\ecor}{\end{cor}}
\newcommand{\bprop}{\begin{prop}}
\newcommand{\eprop}{\end{prop}}
\newcommand{\brem}{\begin{rem}}
\newcommand{\erem}{\end{rem}}
\newcommand{\beq}{\begin{equation}}
\newcommand{\eeq}{\end{equation}}
\newcommand{\beqn}{\begin{eqnarray}}
\newcommand{\eeqn}{\end{eqnarray}}
\newcommand{\beqns}{\begin{eqnarray*}}
\newcommand{\eeqns}{\end{eqnarray*}}

\def \l{\left}
\def \r{\right}
\def \B{\Big}

\newcommand{\BO}{\mathcal{O}}

\newcommand{\nonb}{\nonumber}
\newcommand{\ba}{\begin{array}}
\newcommand{\ea}{\end{array}}
\newcommand{\bi}{\begin{itemize}}
\newcommand{\ei}{\end{itemize}}
\newcommand{\be}{\begin{enumerate}}
\newcommand{\ee}{\end{enumerate}}
\newcommand{\bd}{\begin{description}}
\newcommand{\ed}{\end{description}}
\newcommand{\bal}{\begin{align}}
\newcommand{\bals}{\begin{align*}}
\newcommand{\bs}{\begin{skip}}
\newcommand{\eal}{\end{align}}
\newcommand{\eals}{\end{align*}}
\newcommand{\es}{\end{skip}}
\newcommand{\babs}{\begin{abstract}}
\newcommand{\eabs}{\end{abstract}}

\def\ii{$\mathbf{i}$}

\newcommand{\De}{\Delta}

\newcommand{\ra}{\rightarrow}

\newcommand{\II}{\infty}

\newcommand{\dis}{\displaystyle}
\def \no{\noindent}

\renewcommand{\Pr}{\mathbb{P}}

\def\ii{\mathbf{i}}

\def\bibfmta#1#2#3#4{{#1}, {#2}, \textit{#3}, #4.}
\def\bibfmtb#1#2#3#4{{#1}, \textit{#2}, {#3}, #4.}

\begin{document}

\title{\textbf{Asymptotic Analysis} \\[.1\baselineskip]
\textbf{of a Leader Election Algorithm}}

\author{Christian Lavault\thanks{Corresponding Author:\
C. Lavault\ \ LIPN (UMR CNRS 7030),\ Universit\'e Paris 13,\
99, av. J.-B. Cl\'ement 93430 Villetaneuse,\ France.
Email: lavault@lipn.univ-paris13.fr}
\and
Guy Louchard\thanks{Universit\'e Libre de Bruxelles,
D\'epartement d'Informatique, CP 212, Bd. du Triomphe, B-1050,
Bruxelles, Belgium. Email: louchard@ulb.ac.be}
}
\date{\small \empty}
\maketitle

\begin{abstract}
Itai and Rodeh showed that, on the average, the communication
of a leader election algorithm takes no more than $LN$ bits,
where $L \simeq 2.441716$ and $N$ denotes the size of the ring.
We give a precise asymptotic analysis of the average number
of rounds $M(n)$ required by the algorithm, proving for example
that $\dis M(\infty) := \lim_{n\to \infty} M(n) = 2.441715879\ldots$,
where $n$ is the number of starting candidates in the election.
Accurate asymptotic expressions of the second moment $M^{(2)}(n)$
of the discrete random variable at hand, its probability distribution,
and the generalization to all moments are given. Corresponding
asymptotic expansions $(n\to \infty)$ are provided for sufficiently
large $j$, where $j$ counts the number of rounds. Our numerical
results show that all computations perfectly fit the observed values.
Finally, we investigate the generalization to probability $t/n$,
where $t$ is a non negative real parameter. The real function
$\dis M(\infty,t) := \lim_{n\to \infty} M(n,t)$ is shown to admit
\textit{one unique minimum} $M(\infty,t^{*})$ on the real segment
$(0,2)$. Furthermore, the variations of $M(\infty,t)$ on the whole
real line are also studied in detail.
\end{abstract}

\section{Introduction}
In~\cite{ItRo1,ItRo2}, Itai and Rodeh introduce several symmetry breaking
protocols on rings of size $N$, among which the first is considered here.
They also show that the average communication cost of this particular
leader election algorithm takes no more than $LN$ bits, where the value
of $L$ is computed in~\cite{ItRo2} to be about $2.441716$.

\no However, their method is less direct and less general than the
asymptotic analysis completed in the present paper. Besides, the method
is tailor-made for finding only the average number of rounds required by
the algorithm: the second moment (and \textit{a fortiori} all other moments),
and the probability distribution are not considered in~\cite{ItRo2}.

By contrast, the asymptotic method used in the analysis of our recurrence
relations is very general and quite powerful. All moments as well as the
probability distribution of the random variable can be also mechanically
derived from their asymptotic recurrences. A full asymptotic expansion,
(for large $n$) can be obtained, and it is illustrated for the mean. An
asymptotic approximation of the probability distribution (when $n\to \infty$,
and $j$ gets large enough) is also completed. The latter is derived
by computing singular expansions of generating functions around their
smallest singularity. The present method may serve as a basic brick for
finding the complexity measures of quite a lot of distributed algorithms.

The last Section of the paper is generalizing the problem to
a probability of the form $t/n$, where $t$ is a non negative real
parameter. We show that there exists \textit{one unique} optimal value
$t^{*} = 1.065439\ldots$ on the segment $(0,2)$, where the real
function $M(\infty,t)$ admits \textit{one unique minimum},
$M(\infty,t^{*}) = 2.434810964\ldots$, on the real line. Finally,
the variations of $M(\infty,t)$ when $t > 2$ are investigated in detail.

\subsection{Algorithm scheme and notation}
For the reader's convenience, we rephrase in our own words
the ``symmetry breaking'' (leader election) algorithm designed
in~\cite{ItRo1,ItRo2}.

\smallskip
Consider a ring (cycle) of $N$ indistinguishable processors,
i.e. with no identifiers (the ring is said to be ``symmetric''),
and assume every processor knows $N$.
The leader election algorithm works as follows.

Let $n$ denote the number of \textit{active} processors.
In the first round (initialization), $n = N$ and each processor
is active. At the beginning of each current round, there remains
$1 < n\le N$ active processors along the ring. To compute the
number of \textit{candidates} in the round (i.e. all active
processors that choose to participate in the election), each candidate
sends a pebble. This pebble is passed around the ring, and every
active processor can deduce $n$ by counting the number of pebbles
which passed through. So, in the beginning of a round every active
processor knows $n$ and decides with probability $1/n$ to become
a candidate.

Thus, three cases may happen in a current round:
\begin{itemize}
\def\labelitemi{$\bullet$}

    \item if there is \textit{one} candidate left, it is the leader;

    \item otherwise, the non candidates are rejected (becoming non active),
    and the remaining active processors (the candidates of the current round)
    proceed to the next round of the algorithm;

    \item if \textit{no} active processors chooses to be a candidate,
    all active processors start the next round.
\end{itemize}

\medskip
Throughout the paper, we let $X(n)$ denote the random variable (r.v.)
that counts the number of rounds required to reduce the number
of active processors from $n$ to 1 (choose the leader), when starting
with $n = N$ active processors. The following notations are used.
\begin{eqnarray*}
P(n,j) & := &  \Pr\B(X(n) = j\B),\ \qquad M(n) \;:=\; \mathbb{E}\B(X(n)\B), \cr
& & \cr
M^{(2)}(n) & := & \mathbb{E}\l(X(n)^2\r)\ \qquad \mbox{and}\ %
\qquad \varphi(n) \;:=\; \mathbb{E}\l(e^{-\alpha X(n)}\r).
\end{eqnarray*}
For the sake of simplicity, we also let $M(\infty)$ and
$M^{(2)}(\infty)$ denote $\dis \lim_{n\to \infty} M(n)$ and
$\dis \lim_{n\to \infty} M^{(2)}(n)$ (resp.); similarly, $P(\infty,j)$
denotes $\dis \lim_{n\to \infty} P(n,j)$.

Finally, let $b(n,k)$ denote the probability that $k$ out of $n$ active
processors choose to become candidates, each with probability $1/n$.
In other words,
$$b(n,k) :=\; {n\choose k} \l(\frac{1}{n}\r)^k \l(1-\frac{1}{n}\r)^{n-k}.$$

The recurrence equation for the expectation $M(n)$ is easily derived
from the algorithm scheme.
\begin{equation} \label{eq:basicrec}
M(n) =\; 1 \;+\; \l(1-\frac{1}{n}\r)^n M(n) \;+\; \sum_{k=2}^{n} {n\choose k} %
\l(\frac{1}{n}\r)^k \l(1-\frac{1}{n}\r)^{n-k} M(k)\ \quad \mbox{for}\ \;n > 1,
\end{equation}
and $M(1) = 0$ (by definition).

\section{Asymptotic analysis of the recurrence}
\begin{thm}
The asymptotic average number of rounds required by the algorithm
to elect a leader is the constant\ $ M(\infty)$. When\ $n\to \infty$,
an asymptotic approximation of $M(n)$ writes
\begin{equation} \label{eq:theoremexp}
M(n) \sim\; \frac{1}{1-e^{-1}}\, \l(1 \;+\; \sum_{k\ge 2} %
\frac{e^{-1}}{k!}\, M(k)\r) \;=\; 2.441715879\ldots
\end{equation}

\bigskip \no
The second moment of the discrete r.v. $X(n)$ is asymptotically
$$M^{(2)}(n) \sim\; \frac{1}{1-e^{-1}}\, \l(-1 \;+\; 2M(\infty) %
\;+\; \sum_{k\ge 2} \frac{e^{-1}}{k!}\, M^{(2)}(k)\r) \;=\; 8.794530817\ldots,$$

\no and an asymptotic approximation of its variance ($n\to \infty$) yields
$$\mathrm{var}\B(X(n)\B) \;\sim \;  \frac{1}{(1-e^{-1})^2}\, %
\B( e^{-1} \;+\; (1-e^{-1}) S_2 \;-\; S_1^2 \B) \;=\; 2.832554383\ldots,$$
where $\dis S_1 =\; \sum_{k\ge 2} \frac{e^{-1}}{k!}\, M(k)$\ and
$\dis S_2 =\; \sum_{k\ge 2} \frac{e^{-1}}{k!}\, M^{(2)}(k)$.

\bigskip \no
More generally,
$$\varphi(n) \sim\; \frac{e^{-\alpha}}{1-e^{-(\alpha+1)}} %
\l(e^{-1} \;+\; \sum_{k\ge 2} \frac{e^{-1}}{k!}\, \varphi(k)\r).$$

\bigskip \no
Finally, the probability distribution $P(\infty,j)$ ($n\to \infty$)
satisfies the following asymptotic approximation when $j\to \infty$,
$$P(\infty,j) \;\sim \; \frac{2\rho}{1-2e^{-1}}\;2^{-j},$$
where $\rho = .2950911517\ldots$
\end{thm}

\bigskip
Up until now, we have been unable to use the classical generating
function approach to compute $M(n)$.

\medskip
However, checking that $M(n)$ is bounded is possible. Indeed,
assuming that there exists a positive constant $B(n-1)$ such that
\beq
M(i)\le B(n-1)\ \quad \mbox{for}\ \;i = 1,\ldots, n-1,\ %
\quad \mbox{and}\ \quad B(1) = 0, \label{E21}
\eeq
the following inequality holds
$$M(n) \;\le \; \frac{1}{1-(1-1/n)^n-(1/n)^n}\, %
\l(1 + B(n-1)\, \sum_{k=2}^{n-1} b(n,k)\r).$$
So $M(n)\le B(n)$, with
\begin{equation} \label{eq:reconB}
B(n) \;=\; B(n-1) \;+\; \frac{1-B(n-1)(1-1/n)^{n-1}}{1-(1-1/n)^n-(1/n)^n}\,,
\end{equation}
and $B(1) = 0$. (We show below that $B(n)$ is increasing.)

\medskip Let us first analyze the recurrence~(\ref{eq:reconB}).
If $B(n)$ is converging, it must converge to the fixed point
of Eq.~(\ref{eq:reconB}), i.e. $e$. So, we let
$B(n) = e - \Delta(n)$, and $\Delta(1) = e$.

\medskip \no For fixed $k$ and large $n$,
\begin{eqnarray}
T_n & := & \l(1 - \frac{1}{n}\r)^{n} \;\sim \;e^{-1}\,\l(1 \,-\, %
\frac{1}{2n} \,-\, \frac{5}{24n^2} \,+ \cdots \r) \label{E31} \\
& & \cr \cr
T_{n-k} & := & \l(1 - \frac{1}{n}\r)^{n-k} \sim \;e^{-1}\,\l(1 \,+\, %
\frac{2k-1}{2n} \,+\, \frac{12k^2-5}{24n^2} \,+ \cdots \r). \label{E32}
\end{eqnarray}
We have
\beq
\De(n) =\; a(n)\De(n-1) \;+\; \frac{b(n)}{n}, \label{E4}
\eeq
with
\beqns
a(n) & = & 1 \;-\; \frac{T_{n-1}}{1-T_n-(1/n)^n}, \cr
& & \cr \cr
b(n) & = & n\,\frac{eT_{n-1}\,-\,1}{1-T_n-(1/n)^n}\,.
\eeqns
Note that $n\ge 3$,\ $a(2) = 0$,\ $0 < a(n) < 1/2$,\
and $0 < b(n) < 1$. Several constants will be used in the sequel:
\beqns
c_0 & := & \frac{e-2}{e-1}\,,\ c_1 :=\; \frac12 \frac{e}{e-1}\,,\ c_2 :=\; %
-\frac12 \frac{e-2}{(e-1)^2}\,,\ c_3 :=\; \frac{1}{24} \frac{e(7e-13)}{(e-1)^2}\,, \cr
& & \cr
c_4 & := &\frac{1}{24} \frac{-7e^2+25e-24}{(e-1)^2}\,,\ %
c_5 :=\; c_1 c_2 c_6 + c_3,\ c_6 :=\; \frac{1}{1-c_0}\,,\ c_7 :=\; %
\frac{c_0}{(1-c_0)^2}\,,\ c_8 :=\; c_1 c_7 + c_5 c_6.
\eeqns
For instance, $a(n) \sim c_0 + \BO(1/n)$\ and $b(n) \sim c_1 + \BO(1/n)$.

\no Iterating Eq.~(\ref{E4}) gives
\beqns
\De(n) & = & \prod_{i=0}^{n-2}a(n-i)\De(i) \;+\; %
\sum_{i=0}^{n-2} \frac{b(n-i)}{n-i}\; \prod_{j=0}^{i-1} a(n-j) \cr
& & \cr
& = & \frac1n \sum_{i=0}^{n/2-1} \frac{b(n-i)}{1-i/n} \;\prod_{j=0}^{i-1}a(n-j) %
\;+\; \sum_{i=n/2}^{n-2} \frac{b(n-i)}{n-i} \;\prod_{j=0}^{i-1}a(n-j).
\eeqns
Now,
$$\sum_{i=n/2}^{n-2} \frac{b(n-i)}{n-i} \;\prod_{j=0}^{i-1}a(n-j) %
\;\le \;\frac12 \sum_{i=n/2}^\II (1/2)^i \;\ra \;0\ \qquad (n\ra\II),$$
and so,
$$\De(n)\sim \;c_6 c_1/n.$$
Hence, for $n$ sufficiently large, $\De(n)$ is decreasing, $B(n)$
is increasing and Eq.~(\ref{E21}) holds for $n$.

\no Moreover, $\Delta(n)$ is indeed decreasing to 0 and $B(n)$
converges to $e$.

For the sake of completeness, we can also get a complete
characterization of $\De(n)$.
\beq
\De(n) \sim \;c_0 \De(n-1) \;+\; \frac{c_1 + c_2\De(n-1)}{n} %
\;+\; \frac{c_3 + c_4\De(n-1)}{n^2} \;+\; \BO(1/n^3), \label{E41}
\eeq
proceeding by bootstrapping, we first obtain
$$\De(n)\sim \;c_1 \sum_{i=0}^\II \frac{c_0^i}{n-i} %
\;\sim \;\frac{c_1}{n}\l(c_6 + \frac{c_7}{n}\r),$$
and next, by plugging the above equivalence into Eq.~(\ref{E41}),
$$\De(n)\sim \;\frac{c_1 c_6}{n} \;+\; \frac{c_8}{n^2} \;+\; \BO(1/n^3).$$

\subsection{Asymptotic approximation of $\bm{M(n)}$}
Since $M(n)$ is bounded and positive, the limit can be taken
in~(\ref{eq:basicrec}) for fixed $k$, more generally for $k=o(n^{1/2})$
(see Subsection~\ref{limsum} below). In virtue of Stirling formula
and Eqs.~(\ref{E31})-(\ref{E32}), the summand writes
\begin{equation} \label{eq:Stirlingrecinit}
b(n,k) \sim \; \frac{e^{-1}}{k!} \l(1 \,-\, \frac{k^2 - 3k + 1}{2n} %
\;+\; \frac{3k^4 - 22k^3 + 39k^2 -9k -5}{24n^2} \,+\cdots \r).
\end{equation}
Hence, by Eq.~(\ref{eq:Stirlingrecinit}), the asymptotic
approximation of $M(n)$ is
\begin{equation} \label{eq:Minfty}
M(n) \sim\; \frac{1}{1-e^{-1}}\, \l(1 \;+\; %
\sum_{k\ge 2} \frac{e^{-1}}{k!}\, M(k)\r),
\end{equation}
which is already given in~\cite{ItRo2}.

\medskip
The average number of rounds required by the algorithm follows,
$$M(\infty) =\; \lim_{n\to \infty} M(n) \;=\; 2.441715878809285246587072\ldots$$

\no Numerically, 15 terms are enough to obtain a very good precision:
the error resulting from the sum in Eq.~(\ref{eq:Minfty}) limited
to $\nu$ terms is bounded by
$$\frac{1}{1-e^{-1} }\, \sum_{k>\nu} \frac{1}{k!}\,.$$

\medskip
Note also that if the size of the ring is known to be $N$, the expected bit
complexity of the algorithm is $2.4417158788\ldots N$. It is easily found,
since $N$ bits per round are used on the average in the algorithm.

\brem\ Carrying on with the analysis of $M(n)$ gives
mechanically a complete asymptotic expansion of $M(n)$.
Eqs.~(\ref{eq:basicrec}) and~(\ref{eq:Stirlingrecinit}) lead to\
$M(n) \,\sim \,M(\infty) \,+\, C_1/n \,+\, C_2/n^2 \,+ \cdots$, where
\begin{eqnarray*}
C_1 & = & -\frac{e^{-1}}{2(1-e^{-1})^{2}} \;+\; \sum_{k\ge 2} %
\frac{e^{-1} \B(-k^2 + e^{-1} k^2 + 3k - 3 e^{-1} k - 1 %
+ e^{-1} - e^{-1} \B)}{2(1-e^{-1} )^2 k!}\; M(k) \cr
& & \cr
& = &  -\frac{e^{-1}(1+2e^{-1})}{4(1-e^{-1})^{2}} \;+\; \sum_{k\ge 3} %
\frac{e^{-1} \B((1-e^{-1}) k(3-k) - 1\B)}{2(1-e^{-1})^2 k!}\; M(k) %
\;=\; -.7438715372\ldots
\end{eqnarray*}
The expression of $C_2$ being too long to transcribe, we just give
the result: $C_2 = -.1974635346\ldots$.

\no The convergence of $M(n)$ to $M(\infty)$ is thus very slow:
$\BO\l(n^{-1}\r).$
\erem

\subsection{Interchanging limit and summation} \label{limsum}
There remains to justify the interchange of the limit and the summation
within the sum in Eq.~(\ref{eq:basicrec}), which yields the result
in~(\ref{eq:Minfty}).

\subsubsection{Laplace method}
Since the cutoff point in $b(n,k)$ is approximately $k_0 = n^{1/2}$,
the asymptotic form of the sum $\dis \sum_{2\le k\le n} b(n,k)$\
can be derived from the Laplace method for sums (see \cite{BenOr},
\cite[p.~130-131]{Knuth}), or ``splitting of the sum'' technique.

\bigskip
By taking a suitable positive integer $r = o( n^{1/2})$,
we prove that
\bd
    \item[\textit{i)}]\ the sum $\dis \sum_{k=r}^{n} b(n,k)$
    (the ``right tail'' of the distribution) is small for large $n$, and

    \item[\textit{ii)}]\ $\dis \lim_{n\to \infty} \sum_{k=2}^{r} %
    \l|b(n,k) - \frac{e^{-1}}{k!}\r| \;=\; 0$.
\ed

\vspace{1cm}
\textit{\textbf{i)}}\ The ordinary generating function (OGF) of $b(n,k)$,
$\dis F_n(z) :=\; \sum_{k\ge 0} b(n,k) z^{k}$ is
$$F_n(z) =\; \l(1 - \frac{1-z}{n}\r)^n,$$
and the OGF of $\dis \sum_{r+1\le k\le n} b(n,k)$\ is the
product of $F_n(z)-1$\ and $1/(z-1)$, given by
$$\frac{F_n(z) - 1}{z - 1}\,.$$

\medskip
Considering $\dis \sum_{r\le k\le n} b(n,k)$, Cauchy integral formula yields
$$[z^{r-1}]\,\frac{F_n(z)-1}{z-1} \;=\; \sum_{r\le k\le n} b(n,k) \;=\; %
\frac{1}{2\pi \ii}\int_{\Omega} \frac{F_n(z) - 1}{(z - 1) z^r}\,dz,$$
where $\Omega$ is inside the analyticity domain of the integrand and encircles
the origin. We see that $z = 1$ is not a singularity for the integrand,
so we can neglect the term 1 in the numerator, and asymptotically,
$$\frac{1}{2\pi \ii}\int_{\Omega} \frac{F_n(z) - 1}{(z - 1) z^r}\,dz \;\sim\; %
\frac{1}{2\pi \ii}\int_{\Omega} \frac{\exp\l(n\ln\l(1 - \frac{(1-z)}{n}\r) %
- r\ln(z)\r)}{z - 1}\,dz.$$

\smallskip \no
Again, asymptotically, if we can limit the integration within a neighbourhood
of $z - 1 = o(n)$ (which is checked below), one obtains
$$\frac{1}{2\pi \ii}\int_{\Omega} \frac{\exp\B(-(1-z) - r\ln(z)\B)}{z - 1}\,dz.$$
To equilibrate, we set $z = ry$, which yields
$$\frac{1}{2\pi \ii}\int_{\Omega} \frac{e^{-1}}{ry - 1}\, %
\exp\l(ry-r\B(\ln(y) + \ln(r)\B)\r)\,r\,dy.$$
We now use the Saddle point method. The Saddle point is given by $y^{*} = 1$
(and $z^{*} = r$). So we set $y = 1 + ix$ and, by standard algebra, we obtain
an asymptotic approximation when $n\to \infty$,
$$\sum_{k\geq r} b(n,k) \sim\; \frac{e^{-1} e^{r}}{\sqrt{2\pi}\, r^{r+1/2}\,(1-1/r)},$$
which shows that the right tail of distribution $\sum b(n,k)$
converges indeed to zero when $n\to \infty$.

\bigskip
\textit{\textbf{ii)}}\ Next, from approximation~(\ref{eq:Stirlingrecinit}),
$$\sum_{2\le k\le r} \l|b(n,k) - \frac{e^{-1}}{k!}\r| \;=\; %
O\l(\sum_{2\le k\le r} \frac{e^{-1}}{k!}\,\frac{k^2}{n}\r) %
\;=\; O\l(\frac{r^2}{n}\r),$$
which tends to zero as $n\to \infty$.

\bigskip
Finally, by completing the sum in~(\ref{eq:Minfty}),
it is bounded from above by
$$\sum_{k\geq r} \frac{e^{-1}}{k!},$$
which also tends to zero as $n\to \infty$.

\bigskip
Therefore, interchanging the limit and the summation
in Eq.~(\ref{eq:basicrec}) is proved justified.

\subsubsection{Lebesgue's dominated convergence method}
The latter justification may also use the Lebesgue's
dominated convergence Theorem (see e.g., \cite[p.~27]{Rud}).

\medskip \no
By Stirling formula and Eqs.~(\ref{E31})-(\ref{E32}),
\begin{eqnarray}
\lefteqn{ b(n,k) -\; \frac{e^{-1}}{k!} \; \sim \; \frac{e^{-1}}{k!}\, %
\l( \frac{\exp\B( {\frac{k}{n} - \frac{1}{2n} + \frac{k}{2n^2} + %
\BO\l(\frac{n-k}{n^3}\r)} \B) %
\l(1+\frac{1}{12n}\r)}{e^k \l( 1-\frac{k}{n} \r)^{n-k+1/2} %
\l( 1+\frac{1}{12(n-k)} \r)} \;-\; 1 \r)} \cr
& & \cr \cr
& & \sim \;\frac{e^{-1}}{k!}\, \l( \l( \exp\l( \frac{k(k-3)}{2n} \;-\; %
\sum_{i\ge 2} \frac{k^i}{n^i}\, \frac{2k-i-1}{2i(i+1)} \;+\; \frac{1}{2n} %
\;-\; \frac{k}{2n^2} \;+\; \frac{k/n}{12(1-k/n)} \r) \r)^{-1} - 1 \r). \cr
& & \label{eq:domconv}
\end{eqnarray}
Set $x = k/n$, then
$$b(n,k) -\; \frac{e^{-1}}{k!} \;\sim\; \frac{e^{-1}}{k!}\, \l( \l( \exp\B( nf_1(x) %
\,+\, f_2(x) \,+\, \frac{f_3(x)}{n} \B) \r)^{-1} \,- 1 \r),$$
with
\begin{eqnarray*}
f_1(x) & = & (1-x)\ln(1-x) \;+\; x \;=\; \frac{x^2}{2} \;+\; \BO(x^3), \cr
& & \cr
f_2(x) & = & \frac{1}{2}\,\ln(1-x) \;-\; x \;=\; -\frac{3x}{2} \;+\; \BO(x^2), \cr
& & \cr
f_3(x) & = & \frac{1-x}{2} \;+\; \frac{1}{12}\,\frac{x}{1-x}, \cr
\end{eqnarray*}
and
$$f_1(x)\ge 0,\ f_2(x)\le 0,\ \quad \mbox{for}\ |x|\le 1.$$

\medskip
Thus, for large $n$, the largest root of\
$nf_1(x) + f_2(x) + f_3(x)/n $\ in $[0,1]$ is given by
$$\gamma/n \;+\; \BO\l( n^{-2}\r),$$
with
$$\gamma =\, (3+\sqrt{5})/2 = 2.618033988\ldots,$$
which shows that\ $nf_1(x) + f_2(x) + f_3(x)/n \ge 0$\ for $k\ge 3$
and sufficiently large $n$ (uniformly in $k$).
Checking that it remains true for $k = n - \delta(n)$,
with $\delta(n) = \BO\l(n^{\lambda}\r)$, $\lambda < 1$, is easy.

\medskip
Hence approximation~(\ref{eq:domconv}) is $\le 0$ for large $n$,
and by Lebesgue's dominated convergence Theorem, we can justify
the interchange of the limit and the summation in Eq.~(\ref{eq:basicrec}).

\medskip \no
Note that Eqs.~(\ref{E32}) and (\ref{eq:Stirlingrecinit}) already show
that we must take $k\ge 3$: the coefficient of $1/n$ must be positive.

\subsection{Asymptotic approximation of $\bm{M^{(2)}(n)}$}
We turn now to the computation of $M^{(2)}(n)$.

\bigskip
$M^{(2)}(1) \;=\; 0$, and
\begin{eqnarray*}
M^{(2)}(n) & = & \l(1-\frac{1}{n}\r)^n \mathbb{E}\l((1+X(n))^2\r) %
\;+\; \l(1-\frac{1}{n}\r)^{n-1}\!\cdot 1 \;+\; \sum_{k=2}^{n} b(n,k) %
\mathbb{E}\l((1+X(n))^2\r) \cr
& & \cr
& = & 1 \;+\; 2 \l(1-\frac{1}{n}\r)^n M(n) \;+\; %
\l(1-\frac{1}{n}\r)^n M^{(2)}(n) \cr
& & \cr
& & \qquad \qquad \qquad +\; 2 \sum_{k=2}^{n} b(n,k) M(k) \;+\; %
\sum_{k=2}^{n} b(n,k) M^{(2)}(k).
\end{eqnarray*}

\medskip \no
Hence, when $n\to \infty$ (again, interchanging the operators
may be justified as in Subsection 2.2),
\begin{eqnarray} \label{eq:secondmoment}
M^{(2)}(n) & \sim & \frac{1}{1-e^{-1}}\, \l(1 \;+\; 2e^{-1}M(\infty) %
\;+\; 2\sum_{k\ge 2} \frac{e^{-1}}{k!}\,M(k) \;+\; %
\sum_{k\ge 2} \frac{e^{-1}}{k!}\, M^{(2)}(k) \r) \cr
& & \cr
& \sim & \frac{1}{1-e^{-1}}\, \l(-1 \;+\; 2M(\infty) \;+\; %
\sum_{k\ge 2} \frac{e^{-1}}{k!}\, M^{(2)}(k) \r) \;=\; 8.794530817\ldots
\end{eqnarray}
Of course, a full expansion for large $n$ can also be derived
step by step.

\medskip
Now, since the variance of the r.v. $X(n)$ is defined as
$\mathrm{var}\B(X(n)\B) = M^{(2)}(n) - \B(M(n)\B)^2$,
an asymptotic approximation is straightforward (from Eqs.~(\ref{eq:Minfty})
and (\ref{eq:secondmoment})).
$$\mathrm{var}\B(X(n)\B) \;\sim \;  \frac{1}{(1-e^{-1})^2}\, %
\B( e^{-1} \;+\; (1-e^{-1}) S_2 \;-\; S_1^2 \B) \;=\; 2.832554383\ldots,$$
where $\dis S_1 =\; \sum_{k\ge 2} \frac{e^{-1}}{k!}\, M(k)$\ and
$\dis S_2 =\; \sum_{k\ge 2} \frac{e^{-1}}{k!}\, M^{(2)}(k)$.

\subsection{Generalization}
More generally, using $\dis\varphi(n) = \mathbb{E}\l(e^{-\alpha X(n)}\r)$
as defined in the Introduction,
$$\varphi(n) \;=\; e^{-\alpha}\, \l(\l(1-\frac{1}{n}\r)^n\,\varphi(n) \;+\; %
\l(1-\frac{1}{n}\r)^{n-1}\cdot 1 \;+\; \sum_{k=2}^{n} b(n,k) \varphi(k)\r),$$
with
$$\varphi(1) = 1\ \qquad \mbox{and}\ \qquad \varphi(k) = %
1 \,-\, \alpha M(k) \;+\; \frac{\alpha^{2}}{2}M^{(2)}(k) \;+ \cdots$$

\smallskip \no
Therefore,
$$\varphi(n) \sim\; \frac{e^{-\alpha}}{1-e^{-(\alpha+1)}} %
\l(e^{-1} \;+\; \sum_{k\ge 2} \frac{e^{-1}}{k!}\, \varphi(k)\r).$$

\medskip
Also, from the above relations, all moments asymptotic equations
can mechanically be found.

\bigskip \no
Note that, in contrast to the asymptotic analysis of usual leader
election algorithms (e.g. in~\cite{FiMS,LoPr,Prod}), no periodic
components are arising in the present asymptotic results.

\section{Asymptotic approximation of $\bm{P(n,j)}$}

\subsection{Asymptotic recurrence of $\bm{P(n,j)}$\ ($\bm{n\to \infty}$)}
The following recurrence on $P(n,j)$ stems from Eq.~(\ref{eq:basicrec}).
\begin{eqnarray}
P(n,1) & = & \l(1-\frac{1}{n}\r)^{n-1}, \cr
& & \cr
P(n,j) & = & \l(1-\frac{1}{n}\r)^{n}\,P(n,j-1) \;+\; %
\sum_{k=2}^{n} b(n,k)P(k,j-1)\ \quad \mbox{for}\ \; j > 1. \label{eq:recPnj}
\end{eqnarray}

\medskip \no
And the expression of an asymptotic approximation for large $n$ follows,
\begin{eqnarray}
P(n,1) & \sim & e^{-1}, \cr
P(n,j) & \sim & e^{-1}\,P(\infty,j-1) \;+\; %
\sum_{k\ge 2} \frac{e^{-1}}{k!}\, P(k,j-1)\ %
\quad \mbox{for}\ \; j > 1. \label{eq:asymptrecPnj}
\end{eqnarray}

The above asymptotic approximation on $P(n,j)$ provides the following first
13 values of $P(\infty,j)$ ($j = 1$, \ldots, $13$):
\begin{eqnarray*}
\lefteqn{.3678794411, .2625161028, .1634224110, .0946536614, .0524658088, %
.0282518527, .0149122813,} \cr
& & .0077602315, .0039970064, .0020432067, .0010386252, .0005257697, .0002653262
\end{eqnarray*}

\medskip
\brem
By definition, the following alternative expressions of $M(\infty)$
and $ M^{(2)}(\infty)$ also hold,
$$M(\infty) =\; \sum_{j\ge 1} j P(\infty,j)\ \qquad \mbox{and}\ %
\qquad M^{(2)}(\infty) =\; \sum_{j\ge 1} j^2 P(\infty,j).$$

So, $M(\infty)$\ and $M^{(2)}(\infty)$ could also be computed from
the above definitions. However, more than 15 terms should of course
be required; viz. about 50 terms are actually needed to obtain the
same precision as in the previous computations.
\erem

\subsection{Asymptotic approximation of $\bm{P(\infty,j)}$\ ($\bm{j\to \infty}$)}
Let us now compute an asymptotic approximation for $P(\infty,j)$
when $j$ gets large. First, let
$$D(j) :=\;\sum_{k\ge 2} \frac{e^{-1}}{k!}\, P(k,j).$$
Whence the recurrence relation~(\ref{eq:asymptrecPnj}) also writes
$$P(\infty,j) \;=\; e^{-1} P(\infty,j-1) \;+\; D(j-1).$$

\medskip
Here and in the remainder of the paper, the following ordinary
generating functions (OGF) $H(z)$, $G(z)$ and $\Pi(k,z)$
(of $P(\infty,j)$, $D(j)$ and $P(k,j)$, resp.) are used; we define
\begin{eqnarray}
H(z) & := &  \sum_{j\ge 1} P(\infty,j) z^j,\ \qquad %
G(z) \;:= \; \sum_{j\ge 1} D(j) z^j \qquad \mbox{and} \cr \cr
\Pi(k,z) & := & \sum_{j\ge 1} P(k,j) z^j,\ \; %
\mbox{for any fixed integer}\ k\ge 2. \label{notation}
\end{eqnarray}
 From the OGF $H(z)$ defined in~(\ref{notation}) and
the recurrence~(\ref{eq:asymptrecPnj}), we obtain
$$H(z) \;-\; e^{-1}z \;=\; e^{-1} zH(z) \;+\; zG(z),$$
and
$$H(z) =\; \frac{z\l(G(z) + e^{-1}\r)}{1 - e^{-1}z}\,.$$
So, $H(z)$ has a simple pole at $z = e$.

\medskip \no
Yet, a numerical check in Eq.~(\ref{eq:asymptrecPnj}) shows that
$P(\infty,j) = \Omega\l(e^{-j}\r)$, and thus, $H(z)$ must have
a smaller singularity which is (strictly) less than $e$.

\bigskip
Now, the OGF $\Pi(k,z)$ defined in~(\ref{notation})
and the recurrence relation~(\ref{eq:recPnj}) yield
\begin{equation} \label{eq:recPhi}
\Pi(k,z) \;-\; \l(1-\frac{1}{k}\r)^{k-1}\! z \;=\; %
\l(1-\frac{1}{k}\r)^{k} z\Pi(k,z) %
\;+\; \sum_{\ell=2}^k b(k,\ell) z \Pi(\ell,z),
\end{equation}
which gives, for $k = 2$,
$$\Pi(2,z) \;=\; \frac{z/2}{1-z/2}\,.$$
The above result is of course due to the geometric distribution
of $P(2,j)$, with parameter $1/2$.

\medskip
Hence, $\Pi(2,z)$ has a singularity at $z = 2$, and the singular
expansion of $\Pi(2,z)$ in a domain $\mathcal{D}$ around $z = 2$
stands as
$$\Pi(2,z)\; \asymp \;\frac{1}{1-z/2}\,.$$
Let $\dis R(2) = \lim_{z\to 2}(1-z/2) \Pi(2,z) = 1$. In virtue
of Eq.~(\ref{eq:recPhi}), it is easily seen that $z = 2$ is also
a singularity of all the $\Pi(k,z)$'s for any integer $k\ge 2$.
If we denote
$$R(k) := \; \lim_{z\to 2}(1-z/2) \Pi(k,z),$$
we derive from Eq.~(\ref{eq:recPhi}) that
$$R(k) =\; \l(1-\frac{1}{k}\r)^{k} 2R(k) \;+\; %
\sum_{\ell=2}^k b(k,\ell)\,2R(\ell).$$
When $k$ gets sufficiently large, $R(k)$ can be computed
(15 terms are quite enough for the precision required).

\medskip \no
Since
$$G(z) =\; \sum_{k\ge 2} \frac{e^{-1}}{k!}\; \Pi(k,z),$$
the definition of $\Pi(z)$ in~(\ref{notation}) shows that
$z = 2$ is also a singularity of $G(z)$. By setting
$$\lim_{z\to 2}(1-z/2) \sum_{k\ge 2} \frac{e^{-1}}{k!}\, \Pi(k,z) %
\;=\; \sum_{k\ge 2} \frac{e^{-1}}{k!}\, R(k) \;= \rho = .2950911517\ldots$$
(again, interchanging the sum and the limit may be justified
as in Subsection~\ref{limsum}), the singular expansion of $G(z)$
at $z = 2$ writes
$$G(z) \asymp \;\frac{\rho}{1-z/2}\,.$$

\bigskip
Finally, we obtain the singular expansion of $H(z)$ at $z = 2$,
$$H(z) \asymp \;\frac{2\rho}{(1-2e^{-1})(1-z/2)}\,,$$
and therefore, when $j\to \infty$,
\begin{equation} \label{Pequivalent}
P(\infty,j) \;\sim \; 2.233499118\ldots \;2^{-j}.
\end{equation}

\section{Numerical results}
\begin{figure}[!t]
    \center
    \includegraphics[width=0.45\textwidth,angle=270]{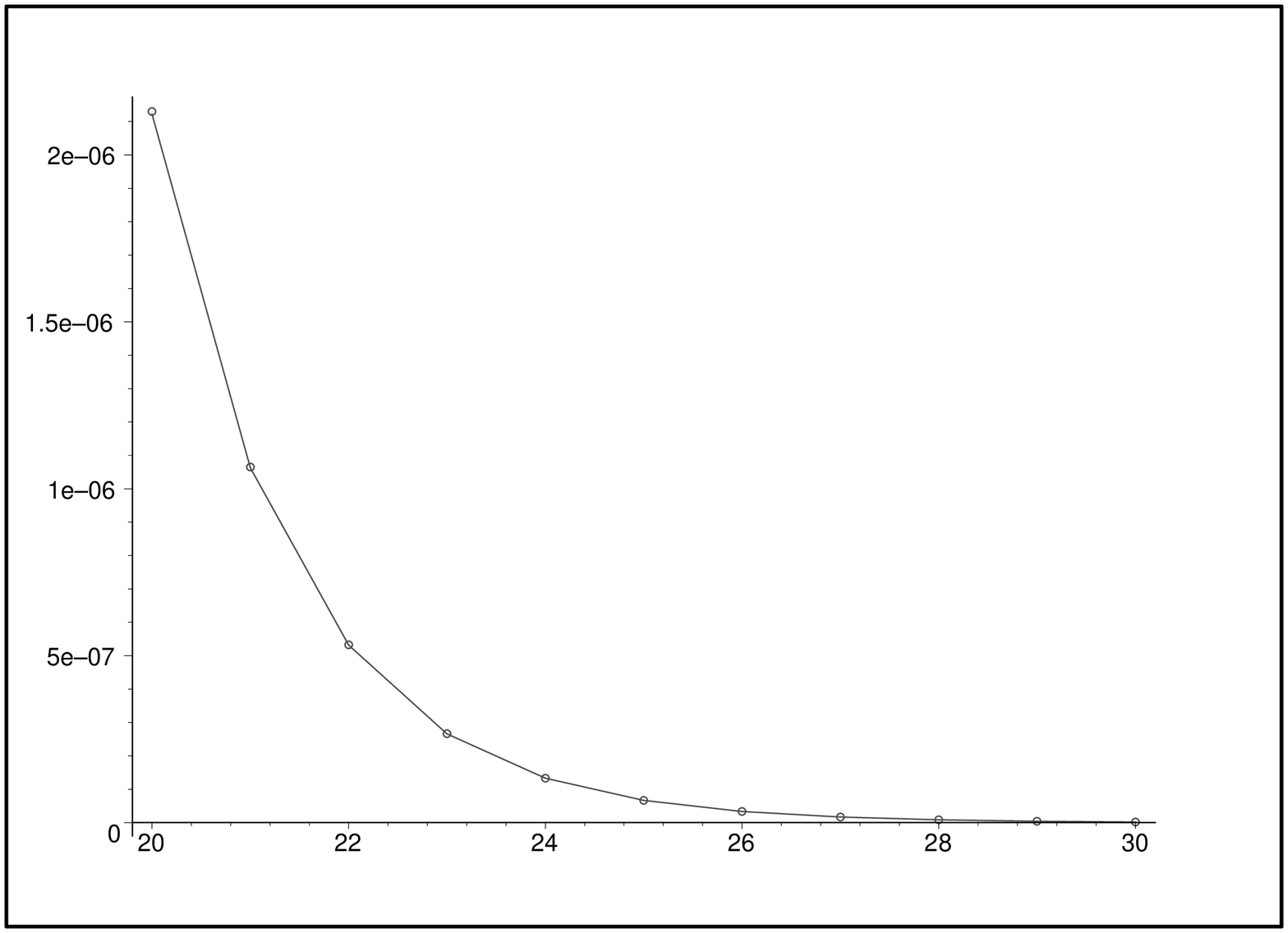}
    \medskip
    \begin{flushleft}
\hspace{3.1cm} $\circ$\, : Observed values of $P(\infty,j)$ \\
\hspace{3cm} --- : Asymptotic approximation of $P(\infty,j)$
in~(\ref{Pequivalent})
\end{flushleft}
     \caption{$P(\infty,j)$ and its asymptotic approximation
in~(\ref{Pequivalent}) for large $j$\ ($j = 20$, \ldots, $30$)}
    \label{F3}
\end{figure}

As can be seen in the following Figures Fig.~\ref{F1} and Fig.~\ref{F2},
the previous computations of $P(\infty,j)$, and $M(\infty)$
and $M^{(2)}(\infty)$ perfectly fit the above ones. Moreover,
Fig.~\ref{F3} shows that the observed values of $P(\infty,j)$
also perfectly fit the asymptotic approximation of $P(\infty,j)$
obtained in~(\ref{Pequivalent}) for sufficiently large $j$.

\newpage
\vspace*{1cm}
\vfill

\begin{figure}[!t]
    \center
    \includegraphics[width=0.45\textwidth,angle=270]{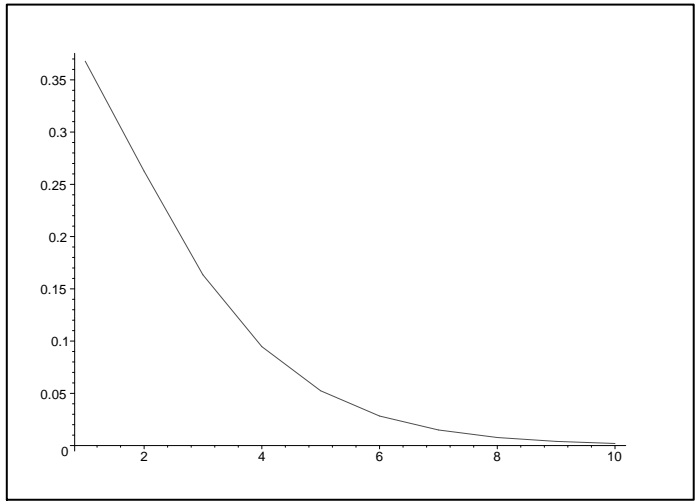}
    \caption{Probability $P(\infty,j)$, for $j = 1$,\ldots, $10$}
    \label{F1}
\end{figure}

\medskip
\begin{figure}[!h]
    \center
    \includegraphics[width=0.45\textwidth,angle=270]{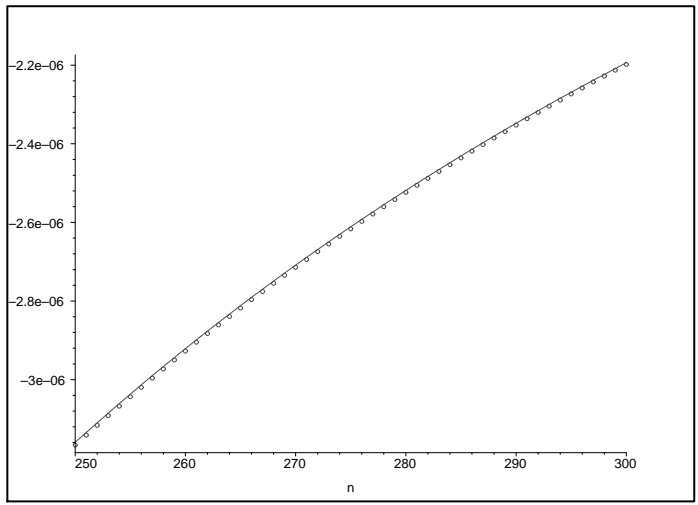}
    \medskip
    \begin{flushleft}
\hspace{3.1cm} $\circ$\, : $M(n) - M(\infty) - C_1/n$ \\
\hspace{3cm} --- : $C_2/n^{2}$
\end{flushleft}
    \caption{Convergence of $M(n)$ to $M(\infty)$, for %
    $n = 250$,\ldots, $300$}
    \label{F2}
\end{figure}
\vfill

\newpage
\vspace*{.05cm}

\section{Is $\bm{1/n}$ the optimal probability?}
Let $t$ be a non negative real number. Following a question
raised by J. Cardinal, let $t/n$ be the probability of choosing
to participate in the election.

\no Is there \textit{one unique optimal} real positive value
$t^{*}$ in some real domain?

\medskip
Taken in the initial context of the first leader election
(``symmetry breaking'') protocol designed in~\cite{ItRo1,ItRo2}
(see Subsection~1.1), $t$ is introduced as a real non negative
parameter which is assumed \textit{known to every processor}
on the ring.

\smallskip
Initially, all the processors are active. At the beginning
of each current round of the election algorithm, every active
processor knows $n$ (the counting process of $n$ is described
in Subsection~1.1), and can decide with probability $t/n$
whether to become a candidate in the round. So, by definition,
$t$ must \textit{a priori} meet the condition $0\le t/n\le 1$.

\bigskip
The recurrence equation for the expectation $M(n,t)$\
(with $0 < t < 2$) is similar to Eq.~(\ref{eq:basicrec}),
\begin{eqnarray} \label{eq:basicrect1}
M(n,t) & = & 1 \;+\; \l(1-\frac{t}{n}\r)^n M(n,t) \cr
& & \cr
& + & \sum_{k=2}^{n} {n\choose k} \l(\frac{t}{n}\r)^k \l(1-\frac{t}{n}\r)^{n-k} M(k,t)\ %
\quad \mbox{for}\ \;n\ge 2, \cr
& & \cr
\mbox{and}\ \ M(1,t) = 0\ \quad  \mbox{(by definition)}. & &
\end{eqnarray}

Upon Differentiating Eq.~(\ref{eq:basicrect1})
with respect to $t$, we obtain
\begin{eqnarray} \label{eq:basicrectdiff}
M'(n,t) & = & - \l(1-\frac{t}{n}\r)^{n-1} M(n,t) %
\;+\; \l(1-\frac{t}{n}\r)^{n} M'(n,t) \nonb \\
& & \cr
& + & \sum_{k=2}^{n} {n\choose k} \frac{k}{n}\, \l(\frac{t}{n}\r)^{k-1} %
\l(1-\frac{t}{n}\r)^{n-k} M(k,t) \nonb \\
& & \cr
& - & \sum_{k=2}^{n} {n\choose k} \l(\frac{t}{n}\r)^k\, \frac{n-k}{n}\, %
\l(1-\frac{t}{n}\r)^{n-k-1} M(k,t) \nonb \\
& & \cr
& + & \sum_{k=2}^{n} {n\choose k} \l(\frac{t}{n}\r)^k %
\l(1-\frac{t}{n}\r)^{n-k} M'(k,t)
\end{eqnarray}
Now, as in Eq.~(\ref{eq:Minfty}), an asymptotic approximation
of $M(n,t)$ for large $n$ yields
\begin{equation} \label{eq:Minftyt}
M(\infty,t) \;=\; 1 \;+\; e^{-t} M(\infty,t) \;+\; %
\sum_{k\ge 2} e^{-t}\, \frac{t^{k}}{k!}\, M(k,t),
\end{equation}
and, similarly, upon differentiating Eq.~(\ref{eq:Minftyt})
with respect to $t$,
\begin{eqnarray*}
M'(\infty,t) & = & - e^{-t} M(\infty,t) \;+\; e^{-t} M'(\infty,t) %
\;+\; \sum_{k\ge 2} e^{-t}\, \frac{t^{k-1}}{(k-1)!}\, M(k,t) \cr
& + & \sum_{k\ge 2} e^{-t}\, \frac{t^{k}}{k!}\, M'(k,t) \;-\; %
\sum_{k\ge 2} e^{-t}\, \frac{t^{k}}{k!}\, M(k,t)
\end{eqnarray*}
or
\begin{equation} \label{eq:Minfytdiff}
M'(\infty,t) \;=\; 1 \;-\; M(\infty,t) \;+\; e^{-t} M'(\infty,t) %
\;+\; \sum_{k\ge 2} e^{-t}\, \frac{t^{k-1}}{(k-1)!}\, M(k,t) %
\;+\; \sum_{k\ge 2} e^{-t}\, \frac{t^{k}}{k!}\, M'(k,t).
\end{equation}

\bigskip \no
Note that the same expression of $M'(\infty,t)$ can also be
derived from the recurrence Eq.~(\ref{eq:basicrectdiff}) by using
asymptotic expansions similar to the ones given in Section 2.

\subsection{Optimal probability on the domain $\bm{(0,2)}$}
A numerical study of the equation $M'(\infty,t) = 0$ on the open
segment $U =\; (0,2)$ easily leads to the solution.

$$t^{*} = 1.0654388051\ldots,\ \quad \mbox{with}\ %
\qquad M(\infty,t^{*}) = 2.4348109638268515517966\ldots$$
The relative gain on $M(\infty,1)$ is a bit larger than
$.0028278945$ (hardly more than $.28$ \%).

\bigskip \no
Since the (necessary) condition $M'(\infty,t^{*}) = 0$
is \textit{not} sufficient for $M(\infty,t)$ to have an extremum
at $t^{*}$, there remains to prove
\be
    \item that $M(\infty,t)$ has a \textit{minimum} at $t^{*}\in U$,

    \item that this minimal solution $t^{*}$ is indeed \textit{unique}
    on the segment $(0,2)$.
\ee
Both results derive from the following Subsection~\ref{convex}.

\subsubsection{$\bm{M(\infty,t)}$ is a strictly convex function
on the segment $\bm{(0,2)}$} \label{convex}
All definitions regarding real and convex functions that are used
in the following may be found in~\cite[Chap.~1 and~3]{Rud}.

\smallskip
Since a \textit{strictly convex} function on some real segment
admits at most \textit{one global minimum} on that segment, both
above results (1 and 2) are shown simultaneously by proving that
$M(\infty,t)$ is indeed a strictly convex positive real function
in $U = (0,2)$.

\bigskip
For the sake of simplicity (and in the line of notations
in Subsection~1.1), we let $M(\infty,t)$ denote
$\dis \lim_{n\to \infty} M(n,t)$,
$$b(n,k;t) :=\; {n\choose k} \l(\frac{t}{n}\r)^k \l(1-\frac{t}{n}\r)^{n-k},$$
and finally, we also use the notation
$$\lambda(n,t) := \frac{1}{1 - (1-t/n)^{n} - (t/n)^{n}}\ %
\qquad \mbox{for}\ \; n\ge 2.$$
Besides, the following form of the basic recurrence
Eq.~(\ref{eq:basicrect1}) is considered:
\begin{equation} \label{eq:basicrect2}
M(n,t) \;=\; \lambda(n,t) \;+\; \lambda(n,t)\, %
\sum_{k=2}^{n-1} b(n,k;t)\, M(k,t)\ \quad \mbox{and}\ \;M(1,t) = 0.
\end{equation}

\medskip
Starting from the above recurrence Eq.~(\ref{eq:basicrect2}),
we show below by induction on $n$, that at any point $t\in U$
and for any integer $n\ge 2$ all functions $M(n,t)$ are
\textit{strictly convex} positive real functions.

\no Therefore, as the pointwise limit of such a sequence
$\B(M(n,t)\B)_{n\ge 2}$\ in $U$, $\dis M(\infty,t) := %
\lim_{n\to \infty} M(n,t)$ will be itself a strictly convex
positive real function in $(0,2)$ (see~\cite[p.~73]{Rud}).

\medskip \no
Note also that, by induction on $n$, all functions $M(n,t)$
$(n\ge 2)$ are in $\mathcal{C}^{\infty}(U,\mathbb{R})$
(i.e., infinitely differentiable in $(0,2)$), and this is also
true for the limit $M(\infty,t)$. In the same line of argument,
$\dis M(n,0^{+}) := \lim_{t\to 0^{+}} M(n,t) = M(n,2^{-}) := %
\lim_{t\to 2^{-}} M(n,t) = +\infty$ for any integer $n\ge 2$,
which remains true in the limit $M(\infty,t)$.

\bigskip
$\bullet$\ \textbf{Basic step.} Whenever $n = 2$, and $n = 3$,
Eq.~(\ref{eq:basicrect1}) yields
$$M(2,t) =\; \frac{2}{t(2-t)}\ \quad \mbox{and}\ \quad %
M(3,t) =\; \frac{18-3t-2t^2}{3t(2-t)(3-t)}\,.$$
So when $k = 2$ and $k = 3$, $M(k,t)$ are two positive functions
in $\mathcal{C}^{\infty}(U,\mathbb{R})$\ s.t.
$M(k,0^{+}) = M(k,2^{-}) = +\infty$.

\no Moreover, since
\begin{eqnarray*}
M''(2,t) & = & 4\;\frac{3t^2-6t+4}{t^3(2-t)^3}\;\ge M''(2,1) = 4\ %
\qquad \mbox{and} \cr
M''(3,t) & = & -2\;\frac{2t^6+9t^5-189t^4+837t^3-1674t^2+1620t-648} %
{3t^3(2-t)^3(3-t)^3}\;> 3,
\end{eqnarray*}
$M(2,t)$ and $M(3,t)$ are two strictly convex functions in $U$.

\bigskip
$\bullet$\ \textbf{Induction Hypothesis.} Assume now that for all
$t\in U$, every function $M(k,t)$ is a \textit{strictly convex}
positive real function in $\mathcal{C}^{\infty}(U,\mathbb{R})$, s.t.
$M(k,0^{+}) = M(k,2^{-}) = +\infty$ for any integer $2\le k < n$.

\smallskip \no
At any point $t\in U$, $\lambda(n,t)\ge 1$ for any positive integer $n$
and $b(n,k;t)\ge 0$ for any pair $(k,n)$ of non negative integers.

\medskip In virtue of Eq.~(\ref{eq:basicrect2}) and the induction
hypothesis, $\dis \lambda(n,t)\, \sum_{k=2}^{n-1} b(n,k;t) M(k,t)$
is a \textit{linear combination} of \textit{strictly convex}
(positive real) functions with \textit{non negative coefficients},
$\lambda(n,t)\times b(n,k;t)$, in $U$.

\no Furthermore, $\lambda(n,t)$ in infinitely differentiable in $U$,
$\dis \lim_{t\to 0^{+}} \lambda(n,t) = +\infty$ and $\lambda(n,2)$
is bounded (except for $n = 2$, since $\lambda(2,2^{-}) = +\infty$).

\bigskip
Next, there remains to prove that $\B(\lambda(n,t)\B)_{n\ge 2}$\
is also a sequence of strictly convex positive real function in $U$.

\medskip \no
For any given $0 < t < 2$ and for any $n\ge 2$, the value $\lambda(n,t)$
enjoys the two following inequalities, which derive from the tight
inequalities shown in~\cite[p.~242]{WhiWat}: for $0\le t/n < 1$,
\begin{equation} \label{eq:inequalities}
\l(1 \;-\; e^{-t}\l(1-\frac{t^{2}}{n}\r) \;-\; \frac{t^{2}}{n^{2}}\r)^{-1} %
\;\le\; \lambda(n,t) \;\le\; \l(1 \;-\; e^{-t} \;-\; \frac{t^{n}}{2^{n}}\r)^{-1}.
\end{equation}
It is easily seen that, for any fixed value of $t\in U$, $\lambda(n,t)$
$(n\ge 2)$ is a strictly increasing sequence, and
$\dis \lim_{n\to \infty} \lambda(n,t) = \frac{1}{1-e^{-t}}\,$.

\smallskip \no On the other hand, $\lambda(n,t)$ is a strictly
decreasing function of $t\in U$ for any fixed $n\ge 2$.

\medskip
In short, since $\lambda''(2,t)\ge 4$ and $\lambda''(3,t)\ge 32/27$,
$\lambda(2,t)$ and $\lambda(3,t)$ are two strictly convex positive
real function in $\mathcal{C}^{\infty}(U,\mathbb{R})$.

\smallskip
Again, the proof is by induction on $n$.
If we assume (\textit{Induction hypothesis}) that, up to any integer
$n\ge 2$, $\lambda(n,t)$ is a strictly convex function of $t$ in $U$,
then $\lambda(n+1,t)$ is indeed a strictly convex function of $t$ on $U$.
For example, assuming that $\lambda''(n,t) > 0$\ for any integer
$n\ge 2$, it is shown after some algebra that $\lambda''(n+1,t)\ge %
\lambda''(n,t) > 0$, by the above two inequalities in Eq.~(\ref{eq:inequalities})
and their resulting properties on $\lambda(n,t)$.

\smallskip
Thus, the positive sequence $\B(\lambda(n,t)\B)_{n\ge 2}$ is
also composed of strictly convex real function in $U$

\bigskip
Finally, Eq.~(\ref{eq:basicrect2}) and the above results show that,
for all $t\in U$ and for any integer $n\ge 2$, $M(n,t)$ is a
\textit{linear combination} of \textit{strictly convex} (positive real)
functions with \textit{non negative coefficients}: $\lambda(n,t)$ and
$\lambda(n,t)\times b(n,k;t)$.

Hence, $\B(M(n,t)\B)_{n\ge 2}$\ is a sequence of \textit{strictly convex}
positive real functions in $\mathcal{C}^{\infty}(U,\mathbb{R})$,
s.t. $M(n,0^{+}) = M(n,2^{-}) = +\infty$.

\begin{figure}[t]
    \center
    \includegraphics[width=0.45\textwidth,angle=270]{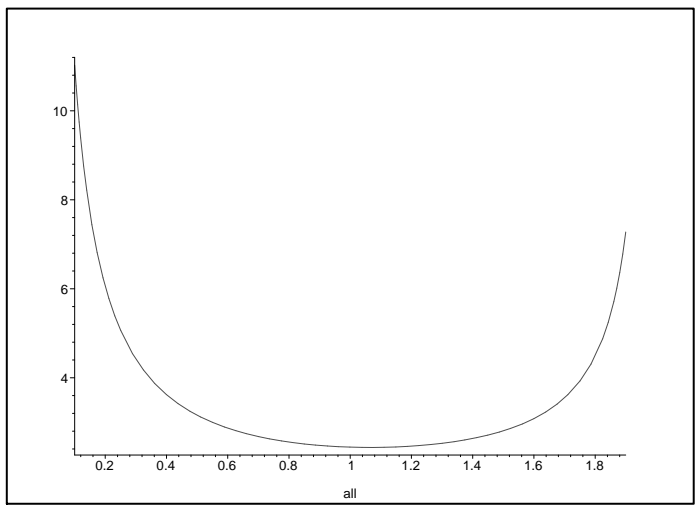}
    \caption{$M(\infty,t),\ t\in (0,2)$} \label{F4}
\end{figure}

\medskip
In conclusion, $M(\infty,t)$ is the pointwise limit of the strictly
increasing sequence $\B(M(n,t)\B)_{n\ge 2}$\ of strictly convex positive
real functions of $t\in U$ (see~\cite[p.~73]{Rud}). Therefore, $M(\infty,t)$
is also \textit{strictly convex} in  $(0,2)$, and the value $M(\infty,t^{*})$
at $t^{*} = 1.065439\ldots$, is \textit{the unique global minimum}
of $M(\infty,t)$ on this segment and we are done. A plot of $M(\infty,t)$
is given in Fig.~\ref{F4}.

\bigskip \no In that sense, we answered the question set in Section~5:
on the real domain $(0,2)$, $t^{*}\!/n$ is indeed \textit{the unique optimal}
probability for an active processor to choose and participate
in the election.

\brem
\vspace*{.5cm}

For any integer $n\ge 2$, $M(n,t)$ is twice differentiable for all
$t\in (0,2)$. Hence, if $M''(n,t) > 0$ the functions $M(n,t)$
are all strictly convex; but the converse is not true.

The positive real function $M(\infty,t)$ is defined on the real
segment $U = (0,2)$ as the pointwise limit of strictly convex
positive real functions defined in $U$. Such is a sufficient condition
for $M(\infty,t)$ to be also strictly convex in $U$.
However the condition is not necessary.

\no Furthermore, $M(\infty,t)$ is the uniform limit of real
functions on any compact subset of the segment $(0,2)$.
This is another way of deriving that sequences of strictly convex
functions do remain strictly convex in the limit on any compact
subinterval of $(0,2)$.
\erem

\section{What happens to $\bm{M(\infty,t)}$ when $\bm{t\ge 2}$?}
There remains to investigate how $M(\infty,t)$ varies
as a function of $t\ge 2$. In the first place, we just assume
that the real parameter $t$ belongs to the domain $(2,3)$.

\subsection{Variation of $\bm{M(\infty,t)}$ in the domain $\bm{[2,3)}$}
Since $t\in (2,3)$ and $0\le t/n\le 1$ (by definition), the value
of the function $M(n,t)$ must be handled separately in the case
when $n = 2$ (i.e. on a ring with two processors).

\no More precisely, two situations may then occur, in which
the \textit{symmetry cannot be broken} with the original algorithm
(see~\cite[p.~1]{ItRo2}:
\bi
    \item if $t = 2$, $M(2,2) = b(2,2;2) = 1$. Both active
    processors on the ring decide with probability $1$ to become
    candidates in each round, and the protocol either perfoms
    an election with \textit{two} leaders, or enters an
    \textit{infinite} computation;

    \item if $2 < t < 3$, we must also set $M(2,t) = 1$
    for the consistency of definitions (when $t\to 3^{-}$,
    $M(2,3^{-}) = +\infty$, as is shown below). In such a case
    \textit{no} termination of the protocol can be achieved.
\ei
Since $M(2,t) = 1$ is set for all $t\in(2,3)$, the recurrence
equation for the expectation $M(n,t)$\ is expressed in a slightly
different form from Eqs.~(\ref{eq:basicrect1}) and (\ref{eq:basicrect2})
on the segment $[2,3)$.
\begin{equation} \label{eq:basicrect3}
M(n,t) \;=\; \lambda(n,t) \;+\; \lambda(n,t)\,b(n,2;t) %
\;+\; \lambda(n,t)\, \sum_{k=3}^{n-1} b(n,k;t)\, M(k,t)\ %
\quad \mbox{and}\ \;M(2,t) = 1,
\end{equation}
where, according to the notation in Subsection~\ref{convex},
$$b(n,2;t) :=\; {n\choose 2} \l(\frac{t}{n}\r)^2 \l(1-\frac{t}{n}\r)^{n-2},\ %
\quad \mbox{and}\ \quad \lambda(n,t) :=\; \frac{1}{1 - (1-t/n)^{n} - (t/n)^{n}}\ %
\quad \mbox{for}\ \; n\ge 3.$$

\bigskip \medskip
There remains to prove that $M'(\infty,t) > 0$\ on the segment
$[2,3)$, with $M(\infty,3^{-}) = +\infty$.

\bigskip
First, following Subsection~\ref{convex} (i.e. again by induction
on $n\ge 3$), $M(n,t)$ in Eq.~(\ref{eq:basicrect3}) is easily shown
to be an increasing sequence of $n\ge 3$ for fixed $t$ in $[2,3)$.

\no Thus, for all $n\ge 3$ and for any $t\in [2,3)$,
$M(n,t)\le M(n,t)\le M(\infty,t)$.

\bigskip
Next, by (modified) Eq.~(\ref{eq:Minftyt}) with $n\ge 3$ and $t\in [2,3)$,
upper and lower bounds on $M(\infty,t)$ are derived.

\no More precisely, after few computations the following
two inequalities hold for all $t\in [2,3)$,
\begin{eqnarray}
M(\infty,t) & \le & \frac{2e^{-t}}{t(t+2)} \;+\; \frac{t}{t+2} \label{in:upM} \\
M(\infty,t) & \ge & \frac{1}{1-e^{-t}}\l( 1 \;+\; \frac{1}{2}t^{2}e^{-t} %
\;+\; M(3,t)\,e^{-t}(e^{t} - t^{2}/2 - t - 1) \r), \label{in:downM}
\end{eqnarray}
where $\dis M(3,t) = \frac{9-3t^{2}-t^{3}}{3t(3-t)}\,$.

\smallskip \no
(Note that since $2.2797\ldots \le M(\infty,2)\le 2.34726\ldots$,
both inequalities (\ref{in:upM}) and (\ref{in:downM}) make sense.)

\bigskip
Finally, Eqs.~(\ref{in:upM}) and (\ref{in:downM}) are used
to bound $M'(\infty,t)$\ from below, and derive that
$M'(\infty,t) > 0$ on the segment $[2,3)$.

\no Indeed, by (modified) Eq.~(\ref{eq:Minfytdiff})
with $t\in [2,3)$, a few calculations yield a lower bound
on $M'(\infty,t)$ for any $t\in [2,3)$.
\begin{equation} \label{eq:lowerboundMprime}
M'(\infty,t) \ge\; \frac{2e^{t}}{t(t+2)} \;-\; %
\frac{2e^{t}(2e^{t}+t^{2})}{t^{2}(t+2)^{2}} \;+\; %
\frac{2(e^{t}-t-1)(9-3t^{2}-t^{3})}{3t(t+2)(3-t)}\,.
\end{equation}
And, since $M'(\infty,t) > 2.26605840\ldots$ for all $t\in [2,3)$,
$M'(\infty,t) > 0$\ on that segment. Furthermore, since all functions
$M(n,t)$\ $(n\ge 3)$\ are in $\mathcal{C}^{\infty}([2,3),\mathbb{R})$\
(see Subsection~\ref{convex}), $M(\infty,3^{-}) = +\infty$ holds
for all $t\in [2,3)$.

\bigskip
Hence, $M(\infty,t)$ is strictly increasing on the segment
$(2,3)$ and $M(\infty,3^{-}) = +\infty$.

\begin{figure}[!t]
    \center
    \includegraphics[width=0.45\textwidth,angle=270]{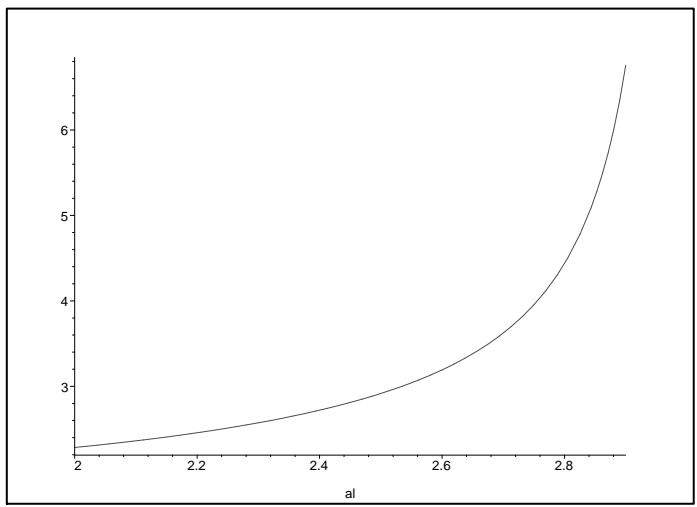}
    \caption{$M(\infty,t),\ t\in [2,3)$} \label{F5}
\vspace*{1cm}
\end{figure}

\smallskip \no
The curve $M(\infty,t)$ is represented in Fig.~\ref{F5}
on the segment $[2,3)$.

\subsection{Variation of $\bm{M(\infty,t)}$ in the domains $\bm{(\xi,\xi+1)}$, %
with $\bm{\xi \ge 3}$}

Investigating the variation of the functions $M(\infty,t)$
when $t\ge 3$ can be carried out along the same lines as in the
previous Subsection 6.1.

\medskip
As can be noticed (e.g. in Subsection~\ref{convex}), the only poles
of the functions $M(n,t)$\ are all the non negative integers 0, 2, 3,\ldots\
(1 excepted) on the real line. Thus, the variation of $M(\infty,t)$
when $t \ge 3$ must be considered on all such consecutive real segments
$(\xi,\xi+1)$, where the $\xi$'s are all integers $\ge 3$.

\medskip
Since $t\in (\xi,\xi+1)$\ still meets the condition $0\le t/n\le 1$\
(by definition), each value $M(\xi,t)$ must again be handled separately
on each open segment $I = (\xi,\xi+1)$.

\smallskip \no More precisely, whenever $n = \xi$ there are $\xi$\
processors on the ring, and the condition $0\le t/\xi \le 1$\
must still hold.
The situation is similar to the one in Subsection 6.1:
the original algorithm \textit{cannot} break the symmetry,
neither if $t = \xi$, nor if $\xi < t < \xi+1$\ (see~\cite[p.~1]{ItRo2}).

\medskip To overcome the difficulty, and for the sake of the consistency
of the definitions, we set $M(\xi,t) := \lceil\lg(\xi)\rceil$\
for all $t\in I$, with $\xi\ge 3$. For example, $M(3,t) := 2$\
(by definition) on the open segment $(3,4)$, and the recurrence
for the expectation $M(n,t)$\ is slightly different from
Eq.~(\ref{eq:basicrect3}) if $t\in (3,4)$.

\no Similarly, each basic recurrence equation for $M(n,t)$
(Eq.~(\ref{eq:basicrect3})), $M'(n,t)$ (Eq.~(\ref{eq:basicrectdiff})),
$M(\infty,t)$ (Eq.~(\ref{eq:Minftyt})) and $M'(\infty,t)$
(Eq.~(\ref{eq:Minfytdiff})) must be adapted to the conditions
on each segment $I$ considered.

\bigskip
On each open segment $I = (\xi,\xi+1)$\ ($\xi\ge 3$),
the variation of the real function $M(\infty,t)$ is roughly
the same. In particular, $M(\infty,t)$ is monotone increasing
in $I$, and it admits no minimum on each such segments.

\section{Conclusions}
As pointed out in the Introduction, performing the asymptotic analyses
of various recurrence relations brings into play some basic, though
powerful, analytic techniques. This is the reason why such methods make
it possible to find easily all moments of the algorithm asymptotic ``cost''
(the numbers of rounds required), especially $M(\infty)$ and $M^{(2)}(\infty)$
(when $n$ gets sufficiently large), as well as an asymptotic approximation
of $P(\infty,j)$ (when $j\to \infty$). The latter is derived by computing
singular expansions of generating functions around their smallest singularity.
Asymptotic expansions of all moments can also be mechanically derived.
All the numerical results performed (with Maple) by both techniques
are quite accurate and fit in perfectly.

Generalizing to a probability $t/n$, where $t$ is a positive real
number, shows that there exists \textit{one unique minimum}
of the function $M(\infty,t)$ on the real segment $(0,2)$~:
$M(\infty,t^{*}) = 2.434810964\ldots$ at the point $t^{*} = 1.065439\ldots$
Besides, the variation of $M(\infty,t)$ whenever $t \ge 2$
shows quite the same behaviour on each real open interval $(\xi,\xi+1)$,
where the $\xi$'s are all the integers $\ge 2$.

In the asymptotic analysis, the major difficulty arises from
the proof of interchanging the limits and the summations
in the recurrences. Two different methods are given that may be
used in many other similar situations:  the Laplace method for sums,
which requires the use of asymptotics via the Saddle point technique,
and the Lebesgue's dominated convergence property.

In conclusion, such analytic techniques may serve as basic bricks
for finding the asymptotic complexity measures of quite a lot of other
algorithms, in distributed or sequential settings.

\section*{Acknowledgements}
We are grateful to both referees, whose very detailed comments
led to improvements in the presentation.

\end{document}